\documentclass[33pt]{article} % use larger type; default would be 30pt

\usepackage[utf8]{inputenc} % set input encoding not needed with XeLaTeX)

\usepackage{geometry} % to change the page dimensions
\usepackage{graphicx} % support the \includegraphics command and options
\usepackage{placeins}
\usepackage{amssymb}
\usepackage{caption}
\usepackage{breqn}
\usepackage{epigraph}
\usepackage{booktabs}
\usepackage{longtable}
\newcommand*{\thead}[1]{%
\multicolumn{1}{c}{\bfseries\begin{tabular}{@{}c@{}}#1\end{tabular}}}

\usepackage{booktabs} % for much better looking tables
\usepackage{array} % for better arrays eg matrices) in maths
\usepackage{paralist} % very flexible & customisable lists eg. enumerate/itemize, etc.)
\usepackage{verbatim} % adds environment for commenting out blocks of text & for better verbatim
\usepackage{subcaption} % make it possible to include more than one captioned figure/table in a single float
\usepackage{lscape}
\usepackage{setspace}
\usepackage{natbib}
\usepackage{fancyhdr} % This should be set AFTER setting up the page geometry
\usepackage{sectsty}
\allsectionsfont{\sffamily\mdseries\upshape} % See the fntguide.pdf for font help)
\usepackage[nottoc,notlof,notlot]{tocbibind} % Put the bibliography in the ToC
\usepackage[titles,subfigure]{tocloft} % Alter the style of the Table of Contents

\usepackage[%  
    colorlinks=true,
    pdfborder={0 0 0},
    linkcolor=blue
]{hyperref}

\hypersetup{
    colorlinks,
    citecolor=blue,
    linkcolor=black
}

\title{Waiting for Dr. Godot: how much and who responds to predicted health care wait times?}
\author{Stephenson Strobel}
\begin{document}
\maketitle

\begin{abstract} 

\noindent Asymmetric information in healthcare implies that patients could have difficulty trading off non-health and health related information. I document effects on patient demand when predicted wait time is disclosed to patients in an emergency department (ED) system. I use a regression discontinuity where EDs with similar predicted wait times display different online wait times to patients. I use impulse response functions estimated by local projections to demonstrate effects of the higher wait time. I find that an additional thirty minutes of wait time results in 15\% fewer waiting patients at urgent cares and 2\% fewer waiting patients at EDs within 3 hours of display. I find that the type of patient that stops using emergency care is triaged as having lower acuity and would have used an urgent care. However, I find that at very high wait times there are declines in all acuity patients including sick patients. 

\phantom{}

\end{abstract}

\newpage
\section{Introduction}
\doublespacing

One of the defining characteristics of healthcare is asymmetric information \citep{arrow_uncertainty_1963}. Interactions between healthcare professionals and patients are designed to transfer knowledge from the professional physician to the naive patient and so improve the patients health. This quality suggests that providing non-health related information to patients could harm their welfare.  For example, in settings where there is an extended wait to see a physician, providing information on those wait times could make sick patients forgo care because they do not fully appreciate how sick they are. Easily interpreted information like wait time is more heavily weighted than not-yet received health information. This justifies keeping that information opaque.
\\

A first step in determining whether information should be withheld from patients is to estimate whether they react in rational ways when that information is disclosed. If very sick patients respond to increased wait times by reducing their visits to emergency settings this provides support for opaqueness in healthcare. Patients do not understand that they are sick and require care. However, if less sick individuals respond while very sick patients do not, this implies that patients have some understanding of their health. \\

Whether patients respond to information provision has implications for health system design. In highly supply constrained environments like emergency departments, provision of properly interpreted information could act as a safety valve on demand. Higher mortality is associated with ED crowding and if less sick patients respond by not seeking care, this could improve outcomes for patients within the ED and hospital \citep{woodworth_swamped_2020, hsuan_association_2023}. This has become more important in settings where poor primary care means patients using ED resources for non-emergent reasons \citep{kelen_emergency_2021, atkinson_saving_2022}.\\

I examine an ED system which began displaying an estimated wait time to patients online and in its ED waiting rooms. This ED system has three advantages. First, because the setting of this study is in Canada, there are no point of care costs in accessing emergency care. This means that the wait time reflects the full ``price" that the patient faces in obtaining care.\\

Second, how estimates of the wait time are created and displayed allows me to causally infer how additional wait time affects patient flows. A machine learning algorithm predicts a new \emph{granular} wait time every six minutes at each ED and these predictions are made to the minute or sub-minute level (ie. a prediction of 37.34 minutes of wait time). However, this ED system publicly displays \emph{coarse} versions of these predictions which are rounded into 30 minute blocks. Thus, a granular wait time of 56 minutes is rounded to a coarse wait time of 30 minutes and a granular wait time of 61 minutes is rounded to a coarse wait time of 60 minutes. This creates a regression discontinuity where EDs with similar granular wait times are assigned different coarse wait times with the latter being observable to patients. I use this RD to construct treatment and control groups of EDs and then estimate impulse response functions by local projections. I assess how effects play out over the three hours after a wait time is displayed.\\

Finally, this ED system has two aspects which allow me to examine the type of patient that are affected by wait time information. This ED system consists of higher resourced ``full" EDs and lower resourced urgent cares (UCs). On the same webpage as the online wait time prediction, the ED system provides guidance on which patients should use each type of site with the goal of streaming lower acuity patients to lower resourced UCs. How demand changes at these different resourced sites likely reflects the relative sickness of the patients that are responding to the wait time information. Similarly, this ED system also assigns an ex ante triage score to each patient prior to any treatment or investigatory decisions. This score is calculated by computer, is relatively objective and reflects how quickly resources should be allocated towards a patient. I use this as a proxy of the immediate health needs of the patient. I can examine whether increased wait times impacts demand by sicker or healthier patients. \\

I find that the overall number of waiting patients in an ED responds to higher wait times. An additional 30 minutes of wait time reduces the overall number of patients at three hours by 0.4 patients. This response varies by ED type; there are declines in the number of waiting patients at both EDs and UCs but there are larger absolute and proportional declines at UCs. I find no overall changes in patients who are graded the highest two levels of urgency (ie. resuscitation and emergent level CTAS) but find declines in the number of patients who are graded as having lower triage scores (ie. urgent, less urgent and non-urgent levels). The response is proportionally highest for patients with sicker triage scores. However this response varies by time; at higher wait times, all types of patients reduce use of the ED.\\

In sum, providing information on wait time to patients causes them to respond in ways that suggest they crudely understand their health. Higher wait times means fewer patients presenting to EDs. The patient who is more likely to respond is one who uses an urgent care and who is triaged as having a lower acuity; these suggest they are in less need of health resources. However, heterogeneity in response by time might imply that displaying wait times that are very high may have an impact on ED demand among very sick patients. \\

\subsection{Previous Literature}

This paper contributes to three major literatures. The first is a set of papers on how informational issues make healthcare a unique market. This dates to an initial literature on the economics of information \citep{stigler_economics_1961} and more specifically on informational inequalities between patients and physicians \citep{arrow_uncertainty_1963}. This spawned a theory based literature examining the effects of information both in and out of the field of health \citep{akerlof_market_1970, spence_job_1973}. \\

This theoretical literature does have empirical support. Having a physician in the family is associated with fewer surgical procedures among family members \citep{domenighetti_revisiting_1993} and can significantly reduce mortality gaps associated with income \citep{chen_roots_2022}. Physician responses to procedure price also suggests patients cannot fully discern whether they require an intervention. Both surgical interventions like Cesarean-sections \citep{gruber_physician_1999} and pap smears \citep{hughes_effect_1992} are more often provided when the physician fee is greater.\\ 

This has led to suggestions that simple pieces of non-health information might have negligible to negative effects on patient behaviour. Healthcare prices are probably the most commonly investigated non-health information transmitted to patients. Price transparency is not a given even with legislation and in the US hospital compliance remains low \citep{jiang_price_2023} with modest effect on prices \citep{han_impact_2022}. In cases where there is effective price transparency, price shopping is not a phenomenon consistently found among health care consumers \citep{brot-goldberg_what_2017, chernew_are_2018}. Suggestions on why this might be include an inability for patients to interpret these information signals properly \citep{milosavljevic_barriers_2023, pollack_necessity_2022} and because insurance inhibits shopping \citep{lieber_does_2017}. Qualitative evidence suggests that trust and the acuity of healthcare needs may trump any consumer impulse to price shop \citep{semigran_patients_2017}. Price transparency can also paradoxically increase the demand for services \citep{kobayashi_impact_2019}. Outside of health, transparency may also cause higher prices through supply sided effects in contexts where perfect competition does not exist \citep{albaek_government-assisted_1997}. Lack of perfect competition often characterizes healthcare although price transparency effects in other settings depend on context and can decrease prices \citep{ater_eects_2022, luco_who_2022}.\\

While prices are a seemingly simple piece of information to convey to consumers they may still be interpreted as an indication of quality. Information on wait times are less subject to this issue. Two medical studies show modest associations between the overall volume of patients presenting to EDs in response to increased wait times \citep{xie_effects_2011, strobel_patients_2021}. Finally, using presentations of high needs patients as an instrument for longer wait times in turn caused lower demand for low-needs patients. The wait time elasticity for these low-acuity patients is estimated as -0.25 \citep{sivey_should_2018}. \\

My contribution to these literatures is three-fold. First, I demonstrate that provision of wait time information does modestly change demand in line with previous medical studies. Second, I demonstrate that there is a negative relationship between resource needs and response; patients that seek care at UCs and those that are less sick are more likely to respond. Finally, I trace out causal effects over several hours using a relatively novel method of identification. The variation in estimated wait times allows me to construct a wait-time demand curve and estimate elasticities for different patient types and estimated wait times. This assessment of heterogeneity suggests that at very high wait times, all types of patients may be affected including the very sick.\\

\section{Methods}

\subsection{Institutional Details}

\begin{figure}
\centering
\includegraphics[width=150mm]{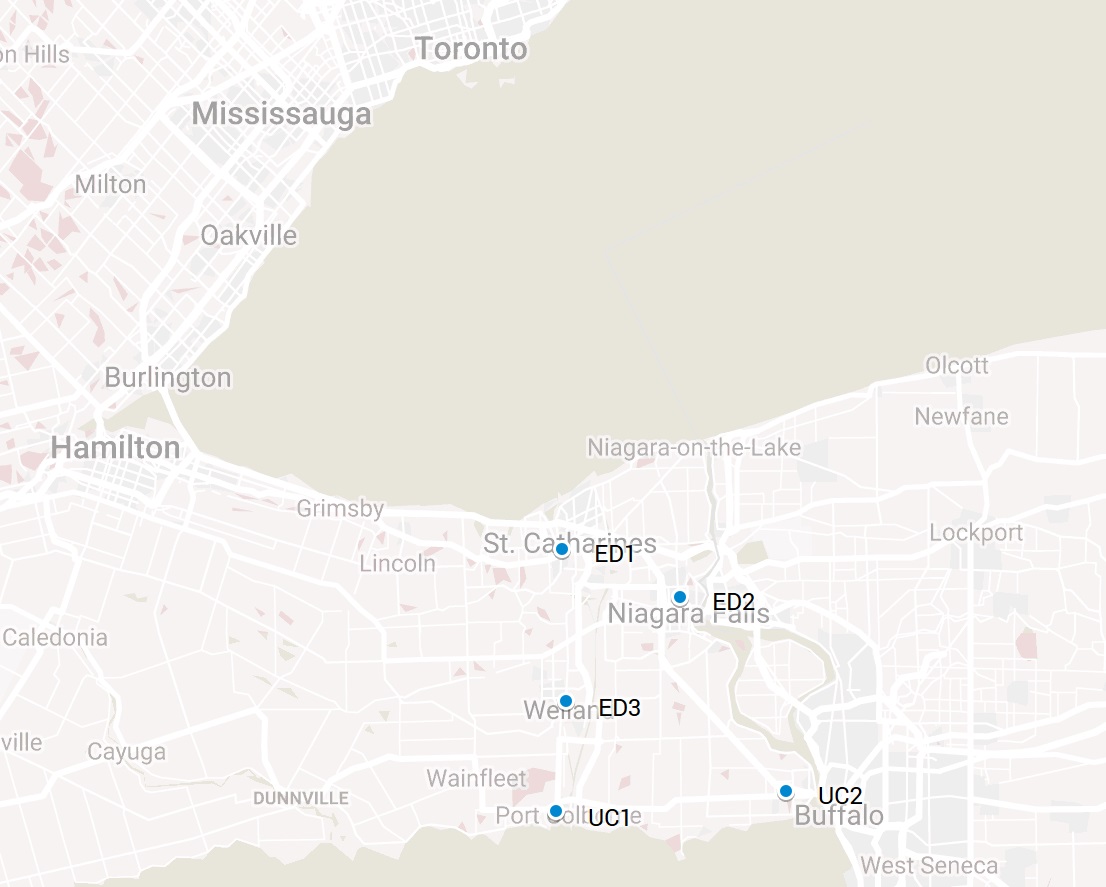}
\caption{A map of the site locations of Niagara Health in southeastern Ontario.}
\label{map}
\end{figure}

This research is set in the Niagara Health system (NHS) ED network, which consists of three ED sites and two UC sites located in southeastern Ontario. The region has a population of approximately 500,000 people, and there are over 200,000 visits to Niagara Health sites every year.  The region is relatively self-contained so that all acute care remains local. To obtain care outside of Niagara Health, a patient would have to cross an international border to obtain care in Buffalo or travel 45 minutes west to the nearest alternate ED system. A map of all sites is displayed in figure \ref{map}.\\

Full EDs are attached to hospitals in St. Catharines (ED1), Niagara Falls (ED2) and Welland (ED3). They have access to 24-hour laboratory services and advanced diagnostic investigations like CT scans and MRIs. There are specialists on call at all sites to admit patients, and to treat patients emergently if necessary.  \\

The two UC sites are at Port Colborne (UC1) and Fort Erie (UC2). They are staffed by ED physicians who also routinely work at the full ED sites and operate 24 hours a day. These urgent care sites have access to x-rays and ultrasounds, but do not have access to consultants, admissions, or advanced imaging (like CT scans) on-site.  If an ED physician deems it necessary to obtain these, the patient must be transferred to one of the full ED sites. UC sites do not routinely accept patients transported by ambulance but any emergent patient presenting to an UC is stabilized and transferred to an ED. \\

\subsection{Details of the wait time predictions}

In 2015 Niagara Health contracted a private company, Oculys Health, to provide services to predict wait times across all ED sites in the system. This was instigated by patient concerns around the uncertainty of ED wait times. Oculys created a proprietary machine learning algorithm that was designed to predict the 90th percentile patients wait time. The prediction algorithm uses real time information from Niagara Health electronic medical records to make new predictions for all five sites every six minutes. These predicted wait times are displayed in each ED waiting room and online for patients to access.\\

\begin{figure}
\centering
\includegraphics[width=150mm]{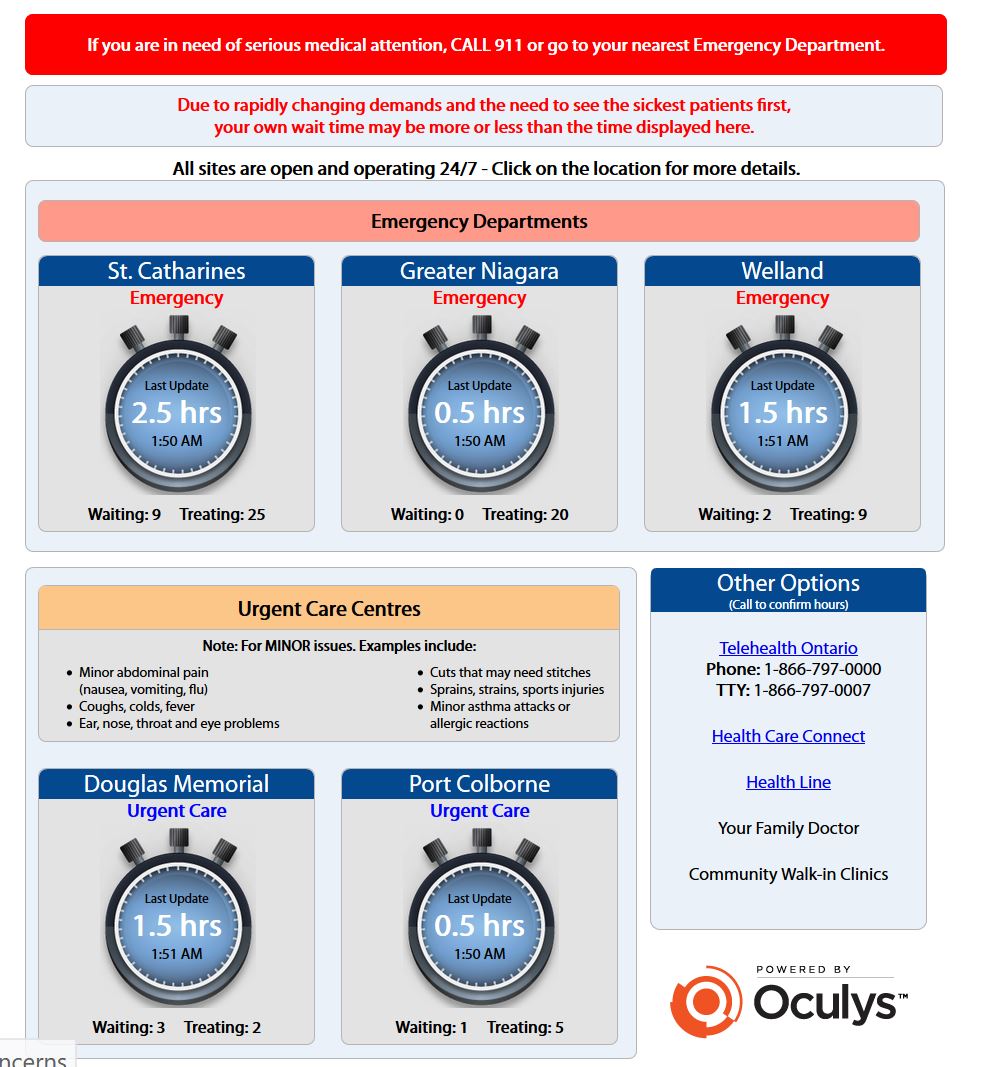}
\caption{An example of wait time predictions that would be displayed online for each site.}
\label{waittime}
\end{figure}

Of importance in causal identification is how the wait time is created and then displayed to patients. The wait time prediction, made by the machine learning algorithm, is at a granular level. This prediction is made at the minute or sub-minute level. These wait times are then rounded to a coarse wait time which is displayed to patients. This rounding is non-intuitive where if a granular wait time is -3 minutes from a 30 minute block, the tracker will round it to that nearest block. For example, if the algorithm predicts a granular wait time between 57 and 86 minutes, the coarse wait time will be displayed online as 60 minutes. If the algorithm predicts a granular wait time between 30 and 56 minutes, the coarse wait time will be displayed online as 30 minutes. Figure \ref{waittime} demonstrates examples of the coarse wait time displayed for patients to view online. In total there are nine RD points where coarse wait times increase by 30 minutes but granular wait times are relatively similar. These range from the 30-60 minute RD to the 270-300 minute RD. Summary statistics for each of these RDs is displayed in table \ref{table summary}.  \\

Niagara Health also instituted additional policies regarding wait time prediction. There is a minimum wait time of 30 minutes posted at all sites and both granular and coarse wait times are censored below this. As it is impossible to know how close a granular wait time is below the cut-off, this RD point is not included in any of the subsequent estimates. The maximum allowable predicted wait time is 300 minutes at EDs and 180 minutes at UCs. However, because I can observe how close the granular wait time is just above the cut-off, these RD points are included. \\

\subsection{Details of the triaging system}

Of additional contextual importance is the triage scoring system that is used across all Niagara Health sites. Since September 2017, all EDs in Ontario have triaged patients using a computerized triage scoring system called the Canadian Triage Acuity Scale (CTAS). CTAS is a discrete score that is assigned to each patient upon presentation to the ED, in order of patient presentation, and before resources are allocated to the patient \citep{beveridge_implementation_1998}. The score ranges from a resuscitation-level case, assigned a score of 1, to a non-urgent case that is assigned a score of 5.  Scores provide guidance on the maximum amount of time a patient should wait before seeing a physician and also correlate with admission probability, suggesting they capture an element of sickness or acuity. Note that \emph{increasing} sickness is correlated with \emph{decreasing} triage score. The descriptions of the tiers are contained in Table \ref{table ctas}. \\

Triaging at these EDs occurs exclusively by nurses who record important variables via a computerized system. These variables which determine the triage score include the patients complaint, their demographics like age, the patients pain score, a set of patient vital signs like their blood pressure, and nursing modifiers which allow nurses to alter the score by one point based on their clinical judgement. This means that the triage score is relatively objective with previous studies confirming high inter-operator reliability \citep{mirhaghi_reliability_2015}.\\

\subsection{Data}

I use two data sets in this research. The first is a data set of predicted wait times for each of the five sites in the ED system; this is a combined dataset of convenience which includes data scraped directly from the wait time website and data provided from Occulys. The difference is that the former data has granular predictions made at the minute level whereas the latter has granular predictions made at the sub-minute level. The scraped data runs from October 2018 to April 2019. Data provided by Oculys runs from October 2017 to September 2018 and August 2019 to November 2019. Predictions are made every six minutes. \\

The second set of data is administrative patient-visit level data from Niagara Health. These contain micro level information on each individual patient visit. Each visit has a CTAS score and date-times associated with when the patient was triaged, the patient was seen by a physician, and when the patient was discharged from the ED. This allows me to construct a minute-by-minute time series of each EDs waiting and treating patient volumes and volume of waiting patients by CTAS level.\\

The main outcome of interest is the stock of waiting patients within the ED. I also examine overall patient treatment volumes. In addition to these primary outcome of interest, I examine the stock of waiting patients by CTAS score.\\

The main variables of interest are whether the ED is treated with an additional 30 minutes of wait time relative to a similar ED. In extensions I am interested in the specific effects of this treatment variable at each site type (ED vs. UC) and at each RD point.\\

\subsection{Identification Strategy}

To examine how wait times change demand for ED care I use impulse response functions estimated by local projection methods \citep{jorda_estimation_2005}. These methods are much more common in macro settings to estimate effects of variables like interest rate changes on other macro outcomes \citep{angrist_causal_2011}. Local projection methods have several advantages, in that they tend to be robust to mis-specification, do not require strict stationarity, and can estimate effects with long forecasts \citep{montiel_olea_local_2021}. This last benefit allows me to robustly estimate effects every five minutes for up to three hours after a wait time is displayed. I estimate a series of regressions for each horizon $h=1,2...,H$

\begin{equation}
    y_{i,t+h}=\psi_{ikh}x_{k,t}+z_t\delta+\mu_{t+h}
\end{equation}

\noindent where $y$ is the stock of waiting patients at an ED site $i$.  A shock to the wait time from exogenous variable $k$ occurs at time $t$ and causes some change to $y$ at $t+h$. In this case, the exogenous shock, $x_{k,t}$ is having a coarse wait time that makes the ED look to have an additional 30 minutes of wait time. This treatment dummy takes a value of one or zero. The parameter of interest is $\psi$ and in macroeconomics is considered the dynamic multiplier coefficient. I include additional controls in $z_t$ which are dummy variables for the hospital site and dummy variables for each individual RD point. When examining the effects of treatment at specific RD points, I interact these latter dummies with the original treatment dummy. Local projections with lag-augmented controls also simplifies the correction of serially correlated standard errors. As such I use Eicker–Huber–White heteroscedasticity-robust standard errors in all regressions \citep{montiel_olea_local_2021}. These regressions are individually estimated for each outcome of interest specified above. \\

I include 14 lags of $y$ and estimate effects up to 36 steps forward. This amounts to lags for 70 minutes of waiting patients and predictions that occur up to 180 minutes after a shock occurs. The former is based on lag-selection procedures that suggest the optimal number of lags are within that range. The latter choice is arbitrary but the geography of the region suggests that any ED site is within an hour drive for a patient. Effects should presumably play out over this time period as a consequence. \\

How is the variable of interest, $x_{k,t}$, assigned to an ED? I leverage the rounding of the granular wait time prediction into a coarse displayed time as a regression discontinuity. At any given point, the granular wait time is predicted to the minute and so ED sites within an arbitrarily small bandwidth are very similar to each other. However, because the granular wait time is rounded to a coarse wait time, similar EDs display different observable wait times to patients. I use this to create treatment and control groups where being just over the cut-off created by the rounding procedure means 30 minutes of additional coarse wait time.\\

\begin{figure}
\centering
\includegraphics[width=150mm]{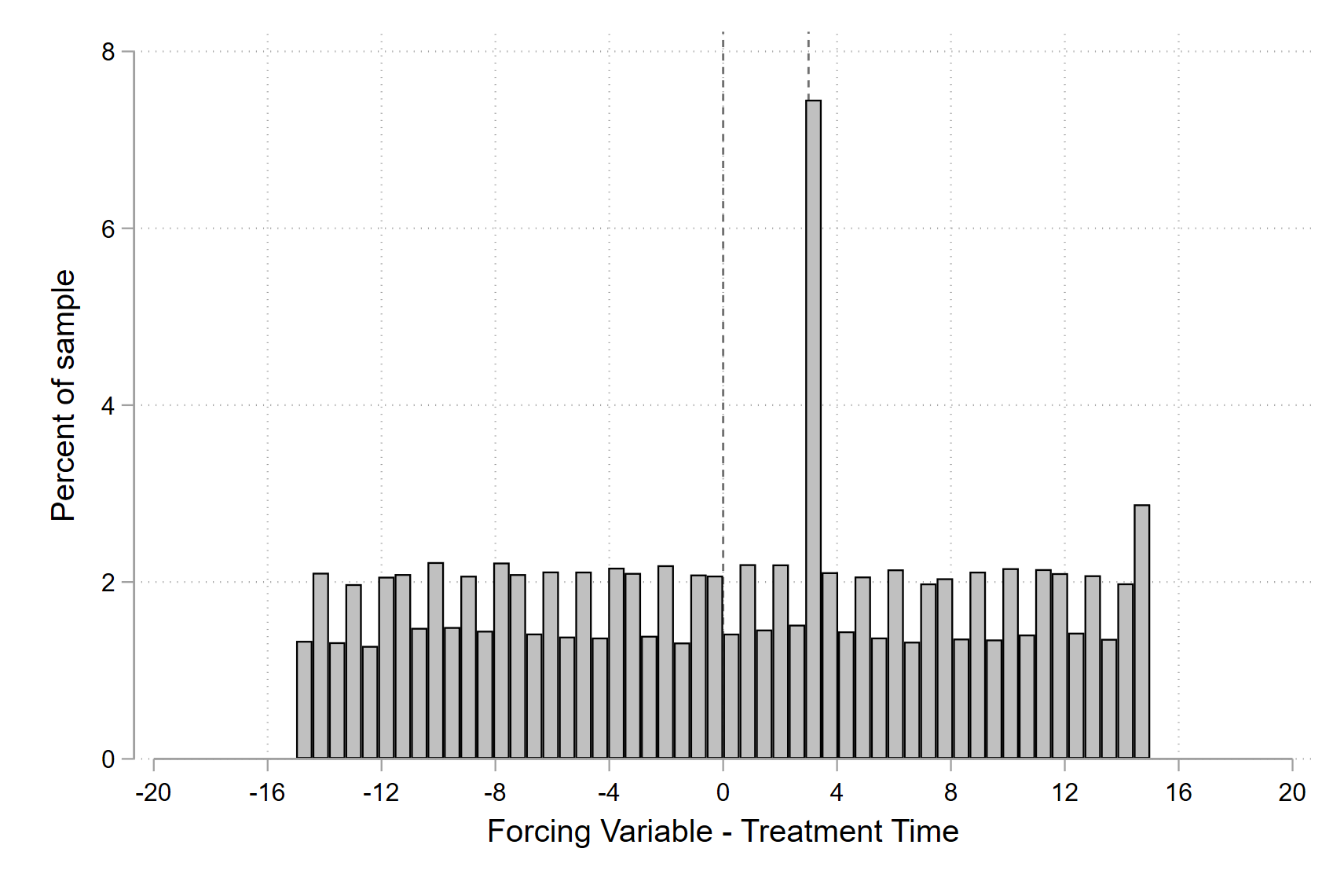}
\caption{The distribution of wait times centered around the cut-off. The forcing variable is the granular wait, centered around zero, which is the point where it is rounded into a coarse wait time.}
\label{mccrary}
\end{figure}

I choose an arbitrary bandwidth that is within 3 minutes of this cut-off. The reason for this is motivated by figure \ref{mccrary}. The wait time prediction data displays two types of ``heaping". Heaping first occurs because this data is partially scraped resulting in increased numbers of observations at each minute of granular wait time. Heaping also occurs because the wait time algorithm preferentially predicts wait times just above the 30 minute mark (ie. it prefers to predict at 60 minutes, 90 minutes etc).\footnote[1]{The reason for this is unclear and private discussions with Oculys staff suggest they were unaware of this occurring.} While not typical of classical manipulation of the running variable it is still concerning \citep{mccrary_manipulation_2008}. To avoid this mass of data I restrict the treatment time to between 0 and 3 minutes around the cut-off and -3 to 0 minutes below the cut-off. As an example, this compares an ED with a granular wait time of 57 to 60 minutes to an ED with a granular wait time of 54 to 57 minutes. \\

\subsection{Calculating wait time elasticities of demand for each RD.}

In my primary results I estimate effects on changes in waiting patients. However, these estimates imply a wait time elasticity of demand \citep{sivey_should_2018}. I calculate

\begin{equation}
\eta_t={({\psi_h \over \mu_{ch}}) \over ({30 \over k_{ch}}) }
\end{equation}

\noindent where the elasticity is the estimated effect, $\psi$ at time $h$, over the mean number of patients at a control site, $\mu_c$. This is divided by the change in coarse wait time of 30 minutes over the baseline wait time at the control sites for that RD $k$. \\

\section{Results}

\begin{figure}
\centering
\includegraphics[width=150mm]{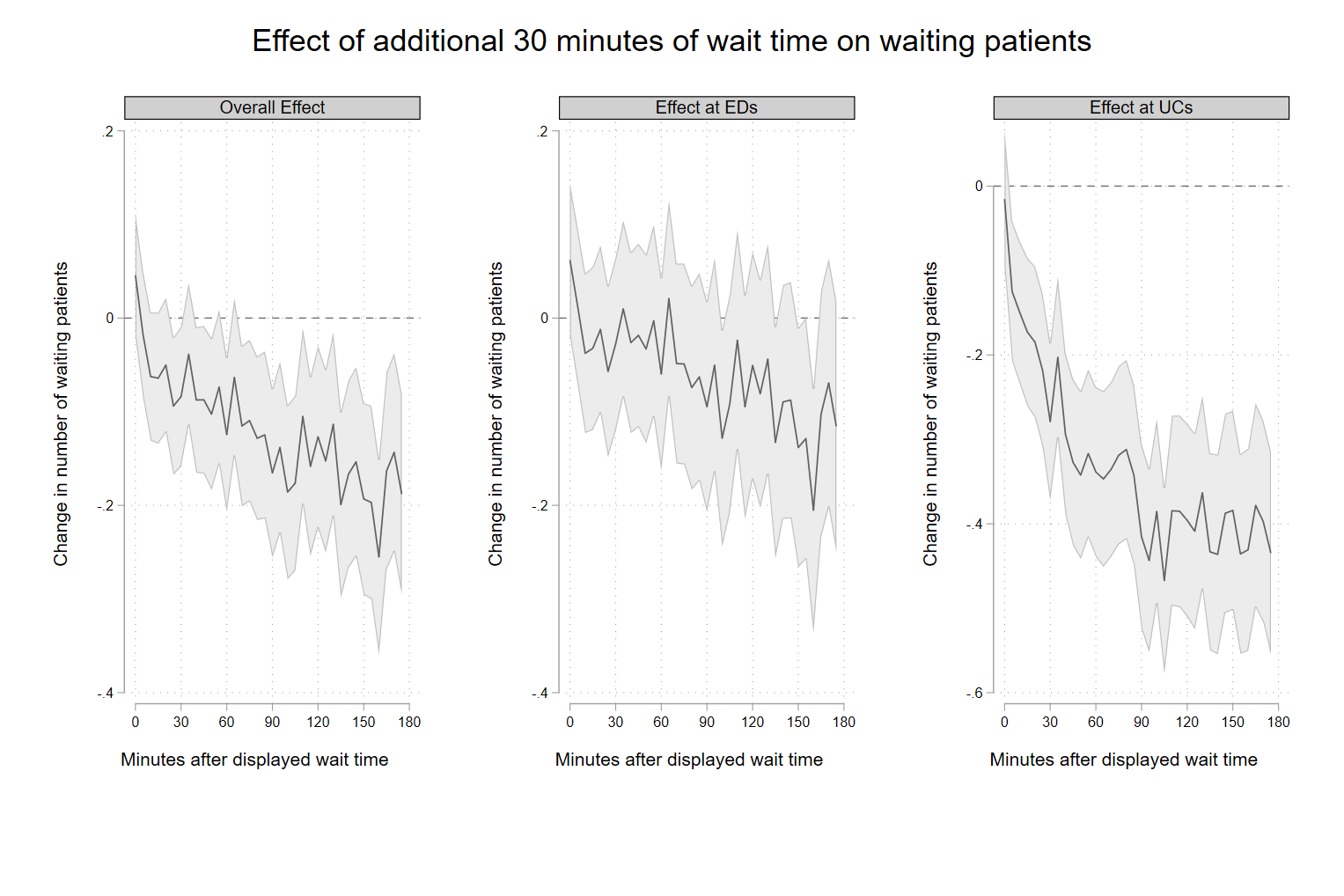}
\caption{Effects of an additional 30 minutes of wait time on the stock of waiting patients for up to three hours after the wait time is posted. Point estimates with 95\% CIs are displayed.}
\label{overall_figure_waiting}
\end{figure}

Figure \ref{overall_figure_waiting} documents the effects on the number of patients waiting in all sites and each ED type for up to 3 hours after a shock occurs. The number of waiting patients declines after a shock by 0.2 at all sites by three hours after a wait time is displayed. This effect plateaus roughly around 90 minutes for all site types. Effects develop more rapidly at UC sites and are of larger magnitude relative to EDs. At UCs there is a 0.4 waiting patient reduction relative to a 0.1 waiting patient reduction three hours after a wait time is displayed.  \\

This change in wait time does not induce physicians to treat more patients. In the case of all sites and ED sites only, the stock of treated patients remains similar up to three hours after a wait time is posted. In the case of UCs though, the effect is to reduce the number of treating patients by about 50\% of the reduction in number of waiting patients. This effect develops after the reduction in waiting patients suggesting that there is some element of patient decompression of the UCs as a result of higher wait times. \\

\begin{figure}
\centering
\includegraphics[width=150mm]{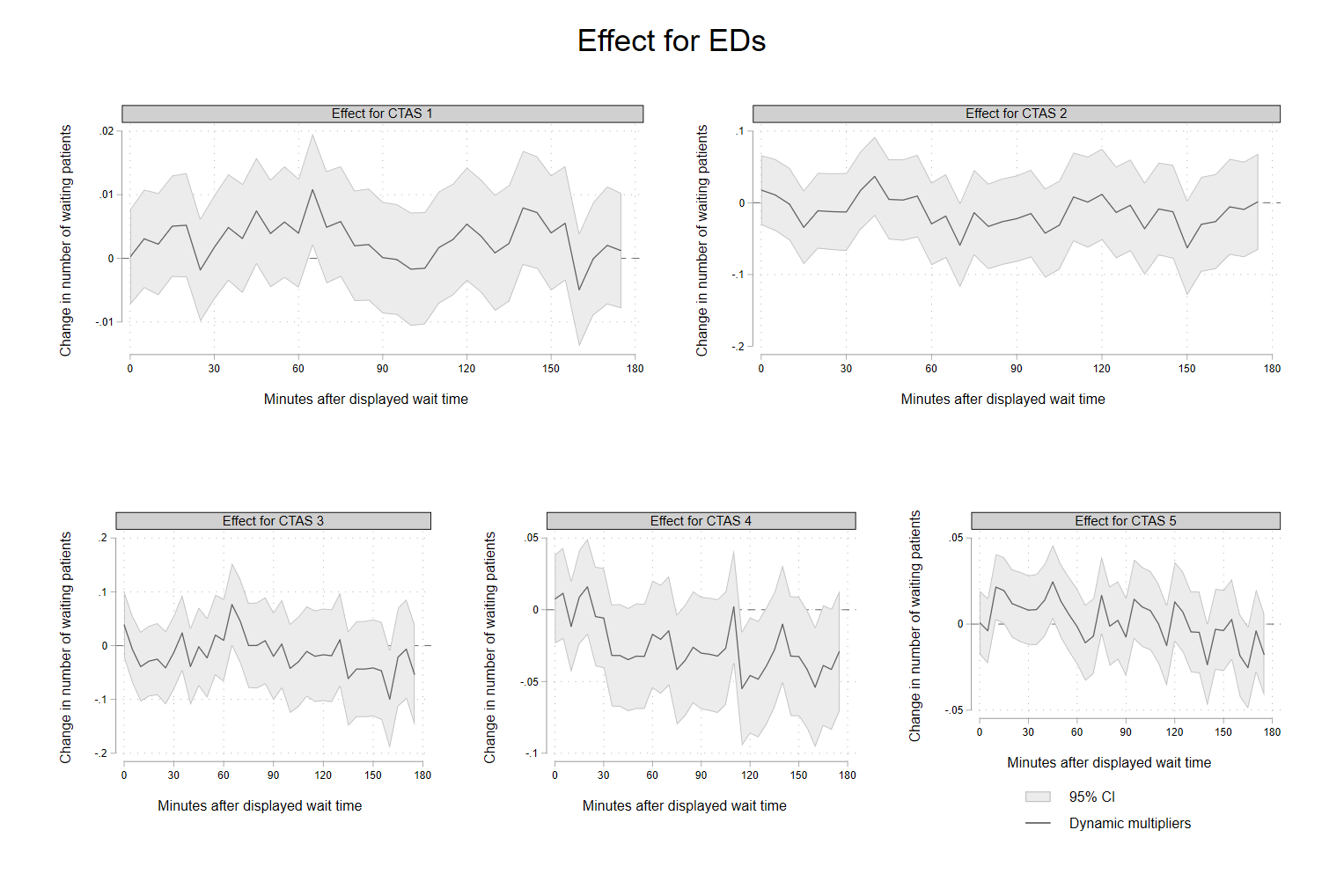}
\caption{Effects of an additional 30 minutes of wait time on the stock of waiting patients at EDs by CTAS for up to three hours after the wait time is posted. Point estimates with 95\% CIs are displayed.}
\label{ed_ctas}
\end{figure}

These effects develop differently by the CTAS level of the patient. At EDs (Figure \ref{ed_ctas}) there are virtually no effects on the number of resuscitation (CTAS 1) or emergency cases (CTAS 2) waiting at the ED for three hours after a wait time is posted. However, there are modest and often statistically noisy reductions in waiting lower acuity CTAS cases at these EDs. This is most apparent in the impulse responses for CTAS 4 patients where there are reductions by 0.05 patients at two hours after a wait time is posted.\\

\begin{figure}
\centering
\includegraphics[width=150mm]{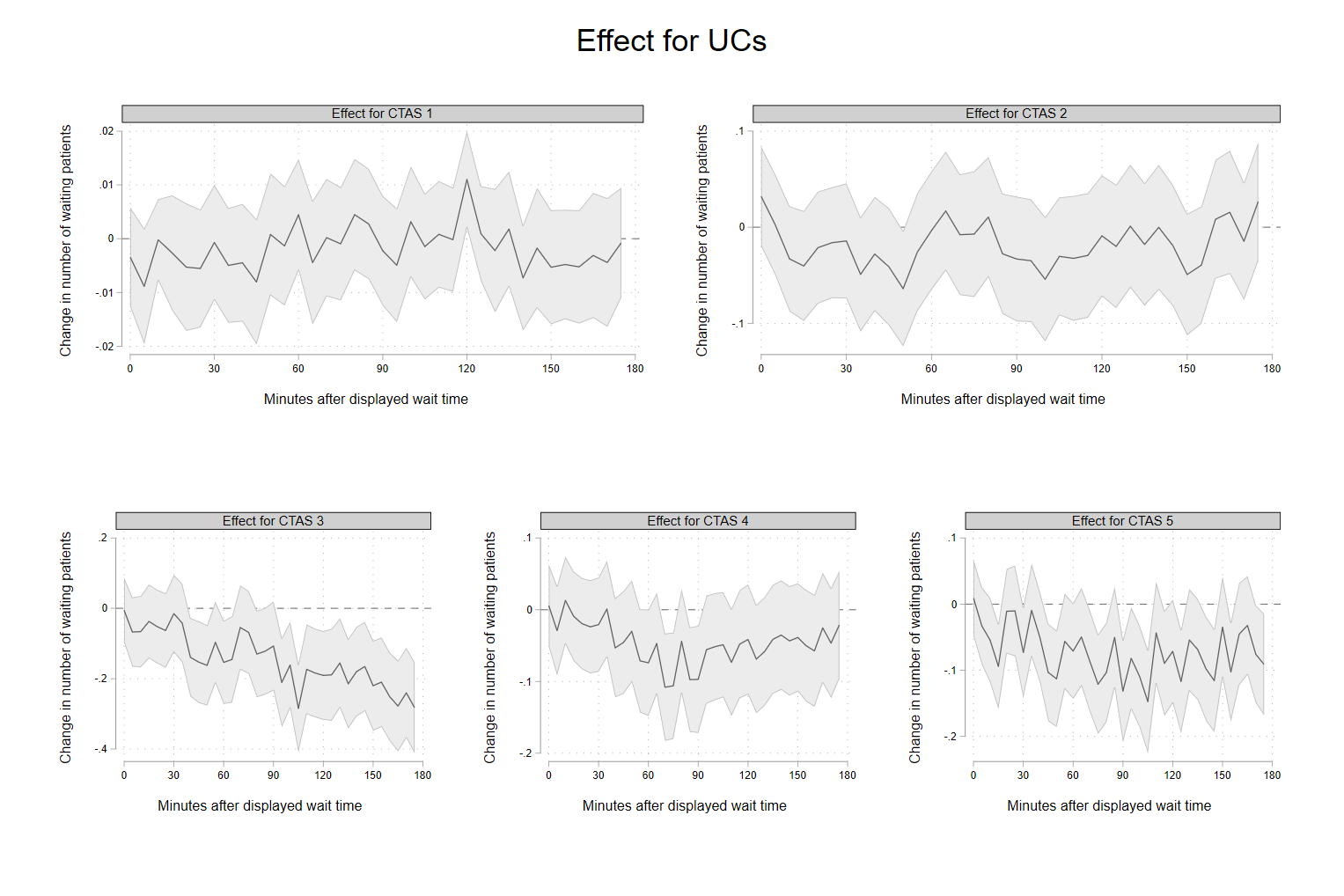}
\caption{Effects of an additional 30 minutes of wait time on the stock of waiting patients at UCs by CTAS for up to three hours after the wait time is posted. Point estimates with 95\% CIs are displayed.}
\label{uc_ctas}
\end{figure}

Effects are most pronounced at UCs but only among lower acuity patients (figure \ref{uc_ctas}). The stock of waiting resuscitation (CTAS 1) and emergency cases (CTAS 2) is not significantly changed. However, there are large and significant declines in the number of waiting CTAS 3, 4, and 5 patients in the three hours after a wait time is posted. This causes a 0.2, 0.1 and 0.1 patient waiting patient decline respectively. \\

\begin{figure}
\centering
\includegraphics[width=150mm]{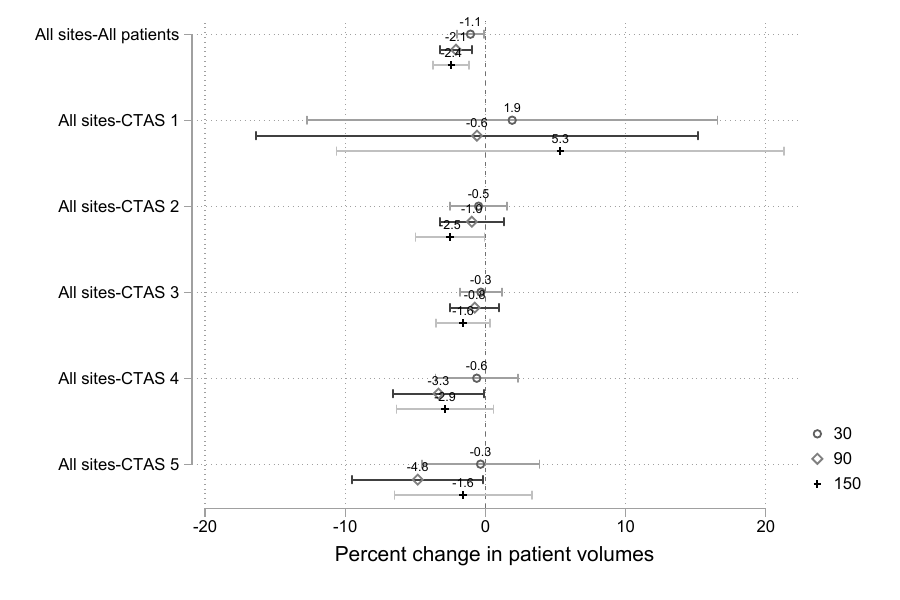}
\caption{Effects of an additional 30 minutes of wait time on the stock of waiting patients at \textbf{all sites} as a percentage of the counterfactual number of patients. Effects are plotted at 30, 90, and 150 minutes after a wait time is posted. Point estimates with 95\% CIs are displayed.}
\label{eclplot_overall}
\end{figure}

However as an overall percentage of the number of waiting patients in the counterfactual control group, lower acuity patients are the most responsive to changes in wait times. Figure \ref{eclplot_overall} demonstrates effects at all sites by overall volume of patients and by CTAS level. The overall effect is to cause a 1.1\% and a 2.4\% decline in total waiting patients at 30 and 150 minutes after a wait time is posted. Effects for each CTAS level are much more noisily estimated, but there are declines by 1.6, 2.9 and 1.6\% of the total volumes of CTAS 3, 4, and 5 patients at 150 minutes. \\

\begin{figure}
\centering
\includegraphics[width=150mm]{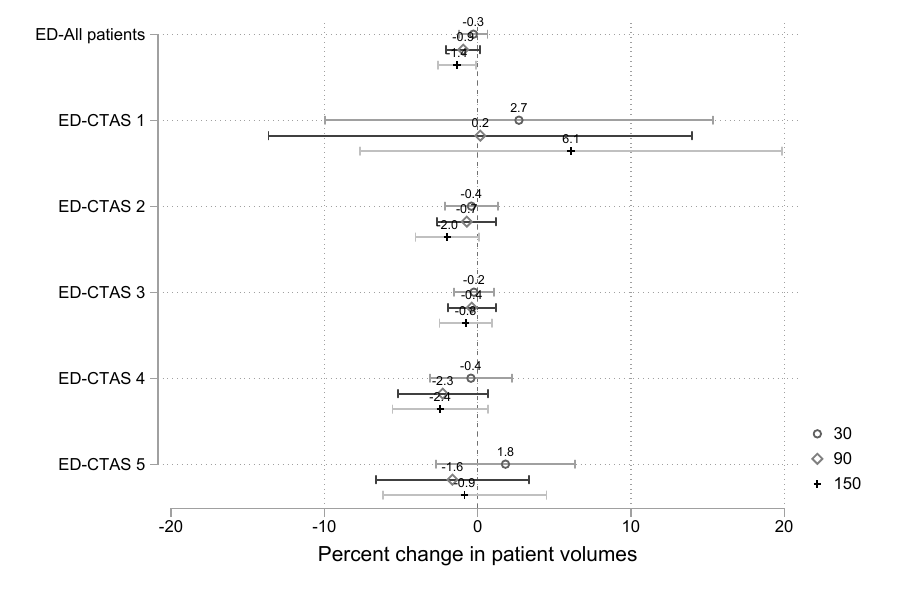}
\caption{Effects of an additional 30 minutes of wait time on the stock of waiting patients at \textbf{EDs} as a percentage of the counterfactual number of patients. Effects are plotted at 30, 90, and 150 minutes after a wait time is posted. Point estimates with 95\% CIs are displayed.}
\label{eclplot_ed}
\end{figure}

\begin{figure}
\centering
\includegraphics[width=150mm]{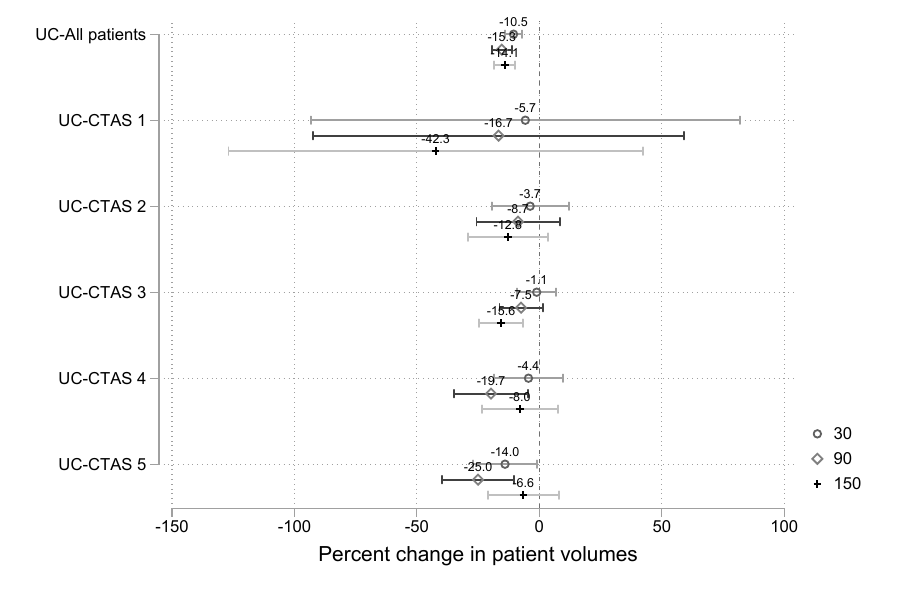}
\caption{Effects of an additional 30 minutes of wait time on the stock of waiting patients at \textbf{UCs} as a percentage of the counterfactual number of patients. Effects are plotted at 30, 90, and 150 minutes after a wait time is posted. Point estimates with 95\% CIs are displayed.}
\label{eclplot_uc}
\end{figure}

This percentage change is also particularly acute at the UCs (figure \ref{eclplot_uc}) as opposed to EDs (figure \ref{eclplot_ed}). Whereas EDs show modest and mostly insignificant declines across the CTAS categories, UCs demonstrate large volume declines in patients. At 90 minutes after a wait time is posted, an additional 30 minutes of wait time results in a reduction in the percentage of overall patients by 15.3\%. This largely affects lower acuity patients with reductions in CTAS 3, 4, and 5 by 7.5\%, 19.7\% and 25\% respectively at 90 minutes after a wait time is displayed.\\

\begin{figure}
\centering
\includegraphics[width=150mm]{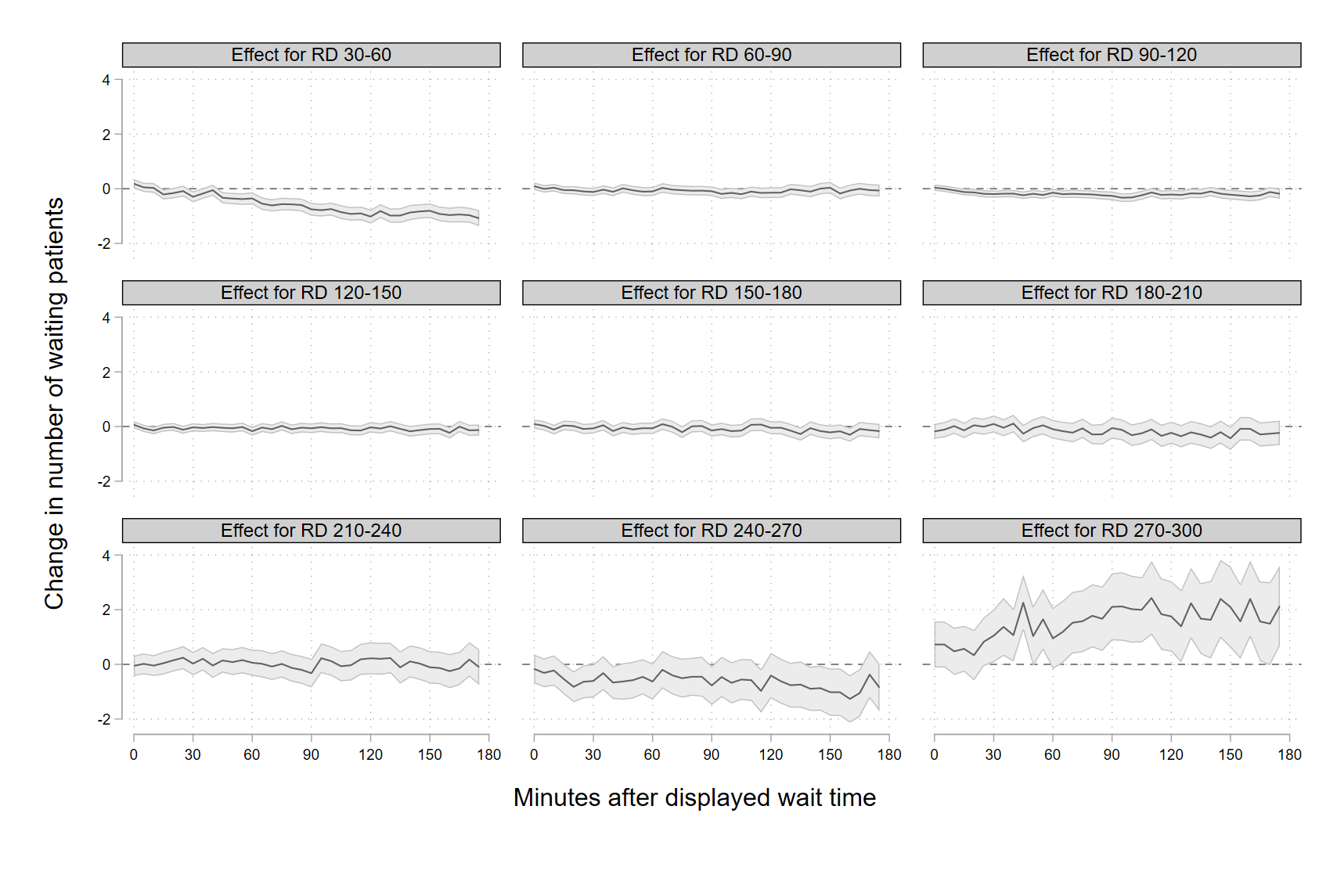}
\caption{Effects of an additional 30 minutes of wait time for each RD.}
\label{rdpoints}
\end{figure}

Finally I examine how the stock of waiting patients evolves by each RD point. Effects across each RD are displayed in figure \ref{rdpoints}. This shows that an additional 30 minutes of wait time results in one fewer waiting patient over three hours at both the 30-60 RD and the 240-270 RD. There are modestly negative to negligible declines for other RDs except for the 270-300 RD. These last estimates suggest that as the wait time increases by 30 minutes the effect is to increase patients over 3 hours. This may be a function of how chaotic an ED is with a wait time of six hours; in these cases triaging may operate with a delay that results in apparent increases of waiting patients after the wait time is displayed.\\

I then estimate elasticities for each of these RDs at 30, 90 and 150 minutes. I remove the RD point of 270-300 because of the above concerns. Given issues of power, I estimate overall effects and effects by patients with low CTAS scores from 3 to 5 and patients with high CTAS scores from 1-2.  \\

\begin{figure}
\centering
\includegraphics[width=150mm]{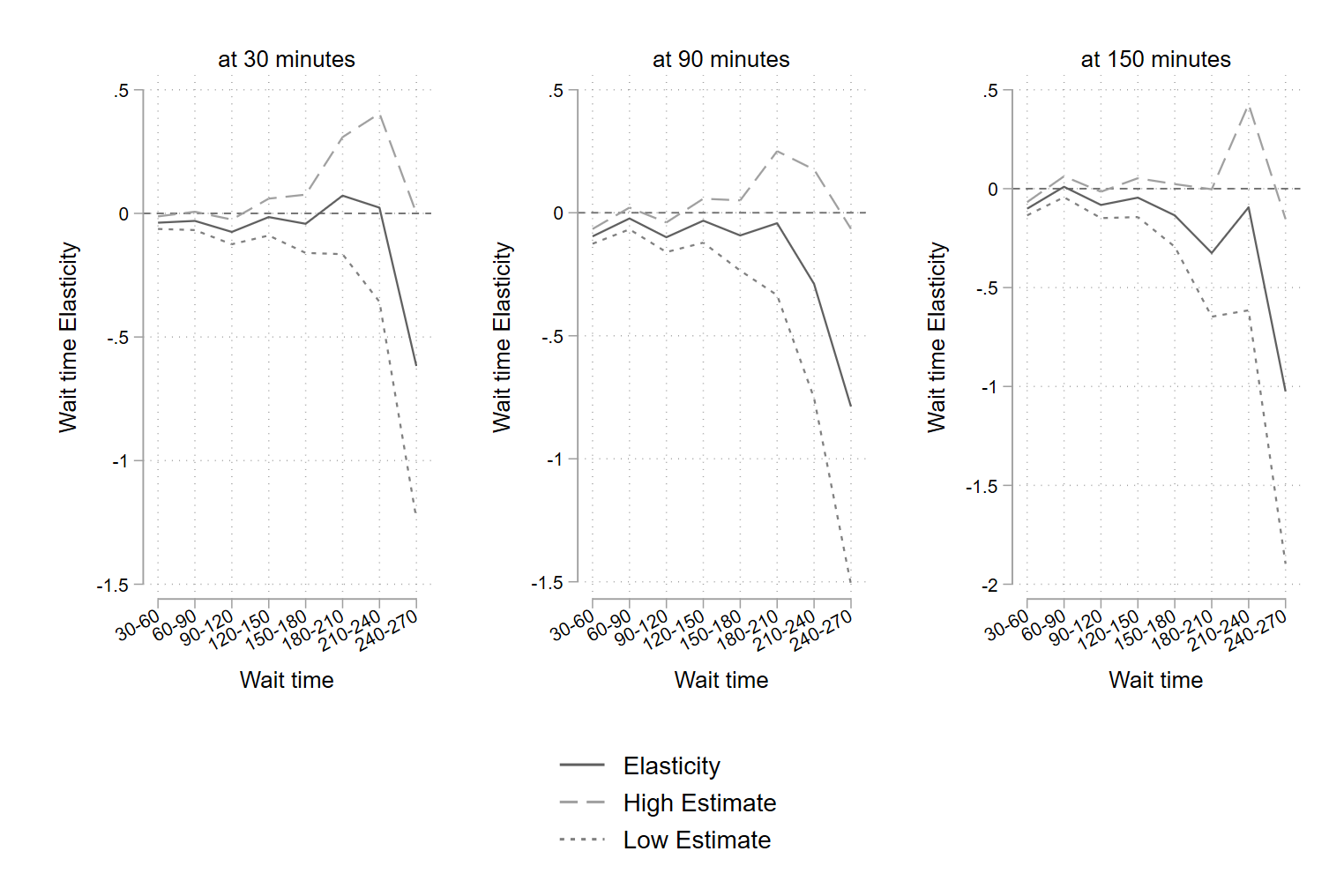}
\caption{Estimated wait time elasticities by RD for all patients. Elasticities are estimate at 30, 90, and 150 minutes after a wait time is displayed. Note that the 270-300 RD is omitted.}
\label{elasticity_all}
\end{figure}

\begin{figure}
\centering
\includegraphics[width=150mm]{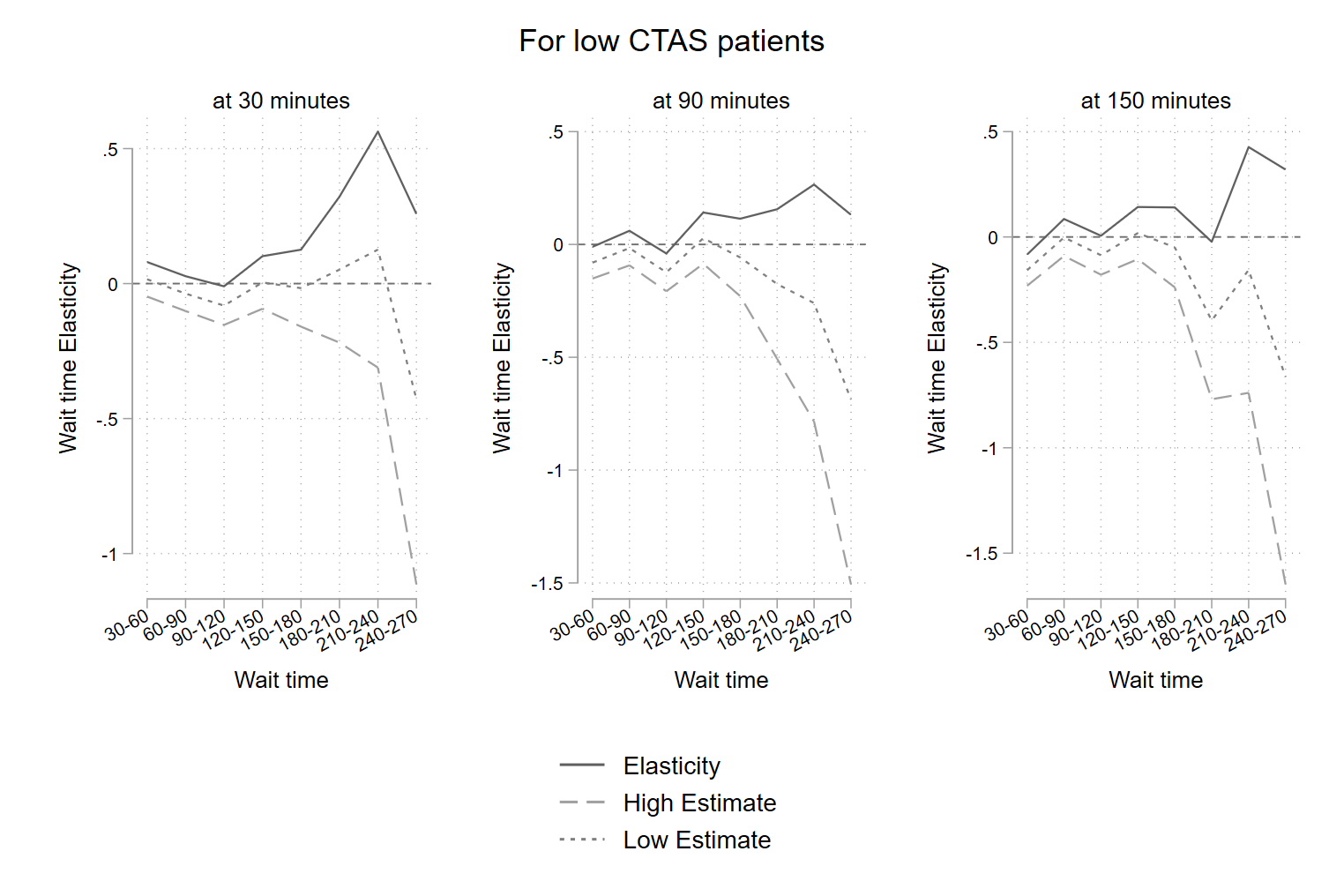}
\caption{Estimated wait time elasticities by RD for low CTAS patients. These are patients who have been triaged as having CTAS 3-5 scores. Elasticities are estimate at 30, 90, and 150 minutes after a wait time is displayed. Note that the 270-300 RD is omitted.}
\label{elasticity_lo}
\end{figure}

\begin{figure}
\centering
\includegraphics[width=150mm]{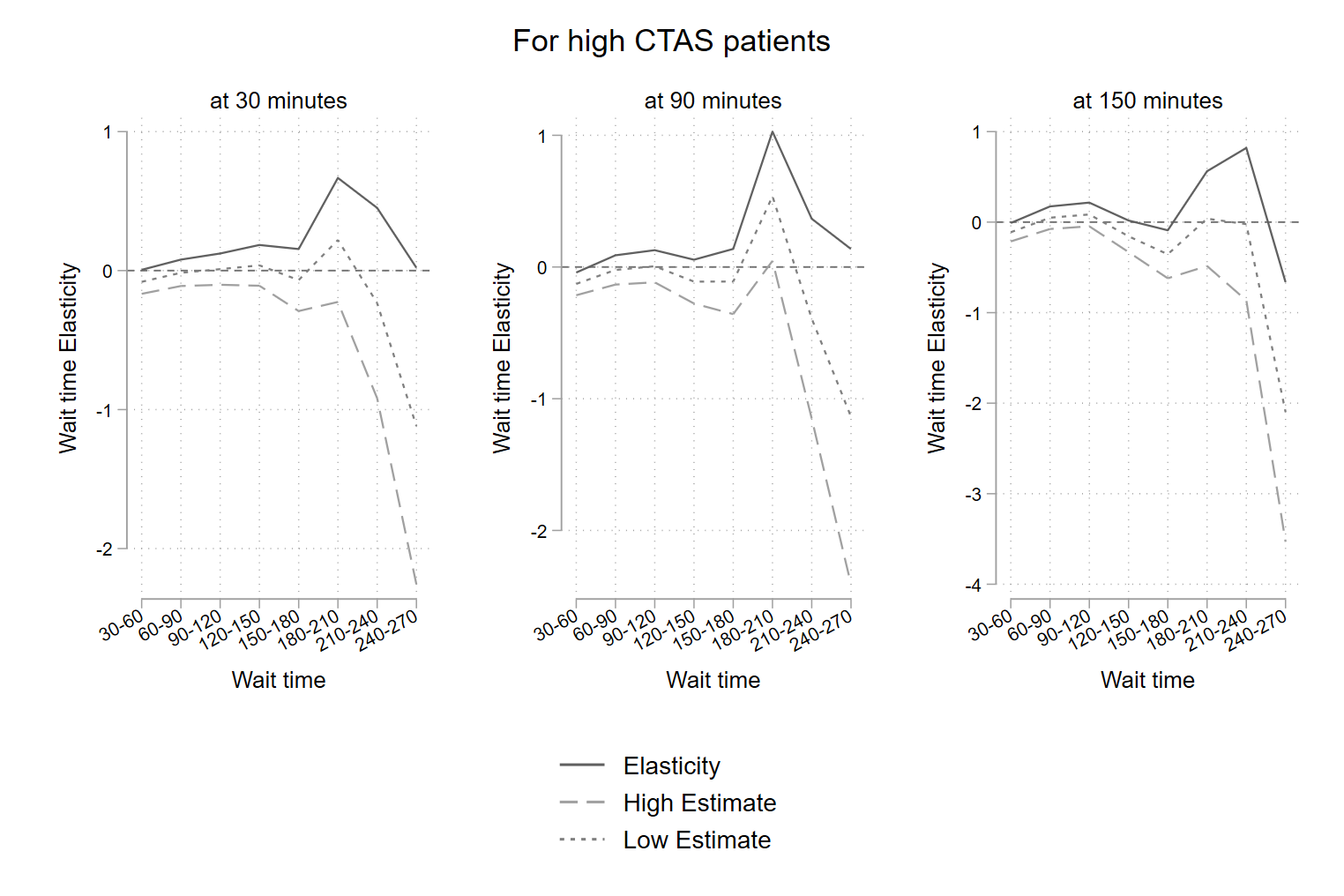}
\caption{Estimated wait time elasticities by RD for high CTAS patients. These are patients who have been triaged as having CTAS 1-2 scores. Elasticities are estimate at 30, 90, and 150 minutes after a wait time is displayed. Note that the 270-300 RD is omitted.}
\label{elasticity_hi}
\end{figure}

Figure \ref{elasticity_all} demonstrates overall effects whereas figure \ref{elasticity_lo} and figure \ref{elasticity_hi} demonstrate effects by low CTAS patients and high CTAS patients. These are noisy estimates but they demonstrate two results. First, there are modestly estimated negative elasticities at lower RDs for the overall sample and the low CTAS patients. At 150 minutes after display, these are in the range of -0.1 to -0.2 for wait times under 210 minutes. For high CTAS patients there are no significant effects at these lower RDs although they are of similar magnitude to low CTAS patients. However as the wait time increases, these negative elasticities increase in magnitude and significance for both low and high CTAS patients. The elasticity for low and high CTAS patients is -0.5 and -2 respectively at 240 minutes of wait time.  \\

\section{Discussion}

Do emergency patients respond to information about the wait they face in obtaining emergency health care? The answer is yes, but modestly so for most wait times. The overall effect is to reduce the stock of waiting patients by 0.2 patients or by 2\% of the overall percentage of waiting patients over 3 hours. Other literature demonstrates similarly modest changes when patients observe changes in their wait times. I estimate the wait time elasticity of demand is -0.1 to -0.2 for most wait times in this data. The first contribution of this paper is to provide additional causal evidence \citep{sivey_should_2018} for these modest effects that have been seen in other  research \citep{xie_effects_2011, strobel_patients_2021}.\\

The second contribution of this paper is to show who decides to stop using ED care. This effect is strongest for patients who would have used lower resourced urgent cares and those patients who are triaged as requiring less urgent care. This former result has been seen in other research where lower acuity sites had larger associated changes in patient volumes with transparent wait times \citep{strobel_patients_2021}.   Both of these outcomes point towards healthier patients as those who are most responsive. This suggests that patients, in a crude sense, have some ability to understand their medical conditions. Truly emergent patients continue to present to the ED even when the wait time is apparently higher. \\

This selected response suggests that there could be scope for policy makers to influence demand for healthcare in a way that affects patients who require care the least. In highly supply constrained environments where mortality may be directly correlated with crowding, this could act as a safety valve on demand swamping care. This seems to be the case at the UCs in this health care system where the stock of treating patients also falls in tandem with the stock of waiting patients. \\

However, this response comes some red flags. At higher wait times, there are larger estimated negative elasticities that suggest that both healthier and sicker patients are being disuaded from using the ED. If the impact of this is to cause declines in welfare because patients cannot understand their underlying health, this could justify keeping this higher wait time information opaque.\\

These results comes with caveats. This paper cannot shed any light on the effects of wait time transparency on mortality or other pertinent health outcomes.  Outside of this, this research has several obvious limitations. It cannot shed light on spillover effects that occur in other settings or outside of this particular ED system. While it would be rare to travel across the border to obtain emergent care it may be the case that patients are being displaced to ED systems west of Niagara Health. Patients may also be visiting family physicians or other primary care settings which would not be captured in this paper. This research has obvious external validity concerns as well being a single ED system in semi-rural Canada.\\

Further research needs to explore several potential avenues. First, as with price transparency, whether there is any effect on patient outcomes like mortality as a result of observable wait times. Second, are there spillover effect onto outpatient care settings. Other studies suggest that retail clinics have the ability to defer ED visits \citep{alexander_check_2019} and the opposite might be the case in this setting.\\

\section{Conclusion}

I demonstrate the impacts of wait time transparency on patient demand for ED care. Patient demand for care responds to higher wait times as evidenced by fewer patients in an ED waiting room for up to three hours after a wait time is displayed. The type of patient responding is one who would have used an urgent care and who is triaged as having a lower acuity presentation. However, at very high wait times there are declines in patient presentations for all acuity levels including very sick patients.\\ 

\newpage
\bibliographystyle{apalike}
\bibliography{Waittimes}{}

\newpage
\section{Tables and Figures}

\begin{table}[h!]{\scriptsize
\begin{center}
\begin{tabular}{ c|c | c | c| c | c | c | c | c | c | c} \hline
&  &\multicolumn{2}{c|}{Observations}  &\multicolumn{2}{c|}{Number of Patients}&\multicolumn{5}{c}{Number of Waiting Patients}\\
 \hline
 Coarse RD & Granular Range & Control & Treatment & Waiting & Treating & CTAS 1 & CTAS 2 & CTAS 3 & CTAS 4 & CTAS 5 \\ \hline
30-60 & 54-60 & 852 & 5089  & 1.22 & 3.126 & 0.01 & 0.2275 & 0.63 & 0.18 & 0.169 \\
60-90 & 84-90 & 2221 & 4729 & 2.15 & 5.887 & 0.0185 & 0.527 & 1.08 & 0.31 & 0.22 \\
90-120 & 114-120 & 3494 & 5430 & 3.36 & 8.05 & 0.023 & 0.897 & 1.70 & 0.476 & 0.26 \\
120-150 & 144-150 & 3558 & 5012 & 5.328 & 10.889 & 0.032 & 1.514 & 2.692 & 0.749 & 0.339\\
150-180 & 174-180 & 2767 & 3015 &8.863 & 15.166 & 0.055 & 2.575 & 4.599 & 1.166 & 0.466\\
180-210 & 204-210 & 1343 & 1418 & 11.772 & 19.72 & 0.082 & 3.62 & 6.032 & 1.529 & 0.507 \\
210-240 &  234-240 & 821 & 940 &13.929 & 20.477 & 0.078 & 4.278 & 7.18 & 1.79 &0.600\\
240-270 & 264-270 & 452 & 443 & 15.037 & 22.07 & 0.0826 & 4.781 & 7.623 & 1.951 & 0.598\\
270-300 & 294-300 & 230 & 156 & 16.35 & 21.96 & 0.106 & 5.16 & 8.417 & 2.03 & 0.634\\
\hline \hline
\end{tabular}
\caption{Summary statistics by RD and key variables of interest.}
\label{table summary}
\end{center}}
\end{table}

\begin{table}[h!]{\scriptsize
\begin{center}
\begin{tabular}{ c|c | c | c| c | c } 
 \hline
 CTAS & Case Description & \thead{\scriptsize Recommended time \\ to see patient} & \thead{\scriptsize Mean time \\ to see patient} & Admission Probability \% & \% of all Cases \\ \hline \hline
1 & Resuscitation & Immediately & 29.02 & 51.3 & 1.53 \\ 
2 & Emergent & 15 minutes & 95.6 & 23.4 & 27.43 \\
3 & Urgent & 30 minutes & 96.4 & 12.3 & 47.57\\
4 & Less Urgent & 60 minutes & 77.3 & 3.1 & 16.93\\
5 & Non-urgent & 120 minutes & 68.9 & 1.5 & 6.53 \\
\hline \hline
\end{tabular}
\caption{CTAS score descriptions}
\label{table ctas}
\end{center}}
\end{table}

\begin{figure}
\centering
\includegraphics[width=150mm]{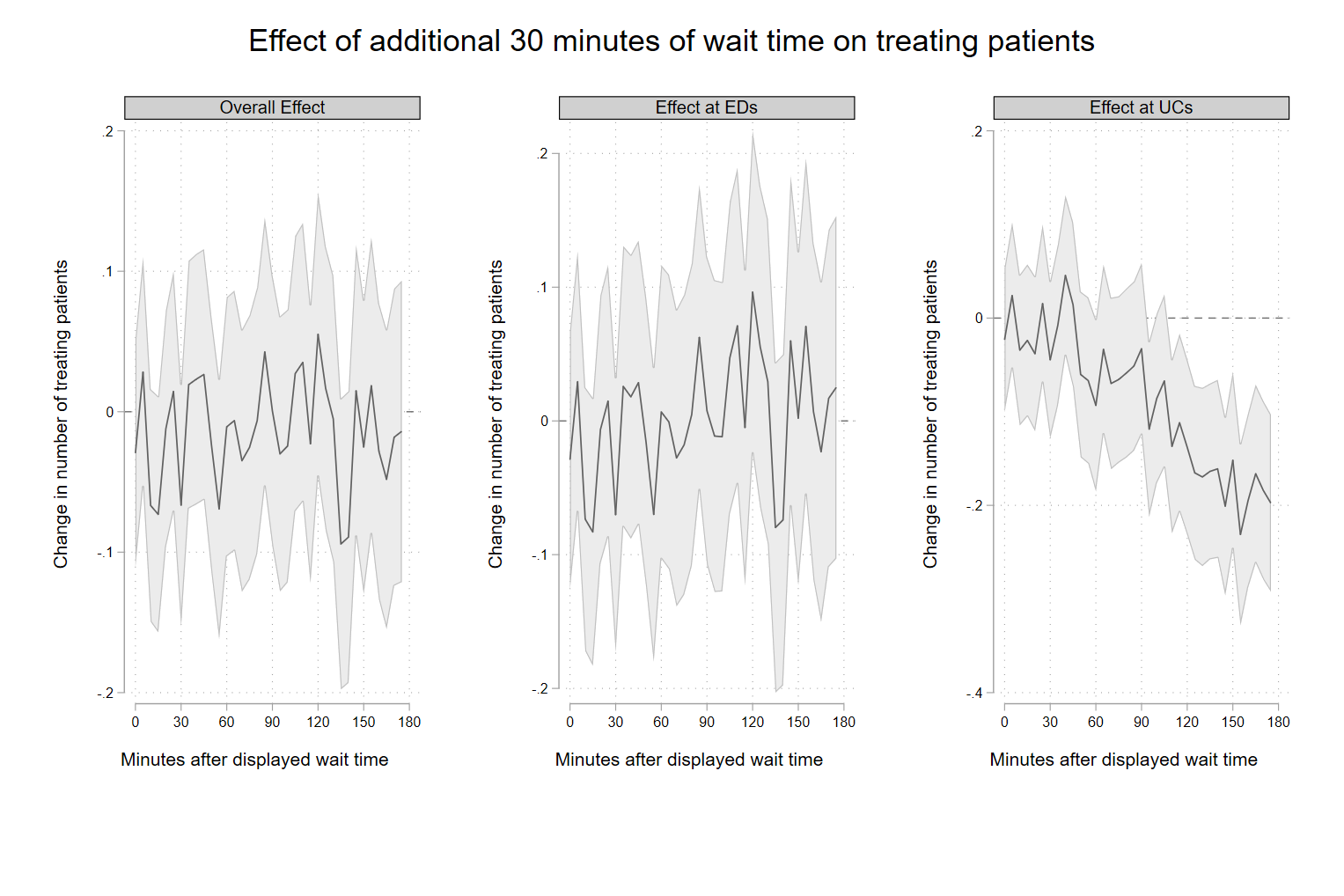}
\caption{Effects of an additional 30 minutes of wait time on the stock of treating patients for up to three hours after the wait time is posted. Point estimates with 95\% CIs are displayed.}
\label{overall_figure_treating}
\end{figure}

\end{document}